\documentclass[%
 reprint,
 superscriptaddress,
 twocolumn, 
 amsmath,amssymb,
 aps,
 pre,
]{revtex4-2}

\usepackage[colorlinks = true,
            linkcolor = blue,
            urlcolor  = cyan,
            citecolor = blue,
            anchorcolor = blue]{hyperref}

\usepackage[T1]{fontenc}
\usepackage[utf8]{inputenc}
\usepackage{graphicx}
\usepackage{dcolumn}
\usepackage{color}
\usepackage{amsmath}
\usepackage[
  color=orange!40,
  textsize=singlespacetiny,
]{todonotes}

\usepackage[charsperline=90]{jlcode}

\global\long\def\mathd{\mathrm{d}}%
\global\long\def\atan{\mathrm{atan}}%

\begin{document}

\preprint{}

\title{A flexible class of exact Hubbard-Stratonovich transformations}

\author{Seher Karakuzu}
\affiliation{Center for Computational Quantum Physics, Flatiron Institute,
162 5th Avenue, New York, NY 10010, USA\looseness=-1}

\author{Benjamin Cohen-Stead}
\affiliation{Department of Physics and Astronomy, The University of Tennessee, Knoxville, TN 37996, USA}
\affiliation{Institute of Advanced Materials and Manufacturing, The University of Tennessee, Knoxville, TN 37996, USA\looseness=-1} 

\author{Cristian D. Batista}
\affiliation{Department of Physics and Astronomy, The University of Tennessee, Knoxville, TN 37996, USA}

\author{Steven Johnston}
\affiliation{Department of Physics and Astronomy, The University of Tennessee, Knoxville, TN 37996, USA}
\affiliation{Institute of Advanced Materials and Manufacturing, The University of Tennessee, Knoxville, TN 37996, USA\looseness=-1} 

\author{Kipton Barros}
\email{kbarros@lanl.gov}
\affiliation{Theoretical Division and CNLS, Los Alamos National Laboratory, Los
Alamos, New Mexico 87545, USA}

\date{\today}

\begin{abstract}
We consider a class of Hubbard-Stratonovich transformations suitable for treating Hubbard interactions in the context of quantum Monte Carlo simulations. A tunable parameter $p$ allows us to continuously vary from a discrete Ising auxiliary field ($p=\infty$) to a compact auxiliary field that couples to electrons sinusoidally ($p=0$). In tests on the single-band square and triangular Hubbard models, we find that the severity of the sign problem decreases systematically with increasing $p$. Selecting $p$ finite, however, enables continuous sampling methods like the Langevin or Hamiltonian Monte Carlo methods. We explore the tradeoffs between various simulation methods through numerical benchmarks.
\end{abstract}

\maketitle

\section{Introduction}

This paper develops a class of Hubbard-Stratonovich (HS) transformations that can be used to handle on-site Hubbard interactions, $\hat{\mathcal{H}}_{\mathrm{int}}=U\hat{n}_{\uparrow}\hat{n}_{\downarrow}$,
in the context of determinant quantum Monte Carlo (DQMC) and related simulation methods~\cite{Blankenbecler81, WhitePRB1989, Gubernatis16}.
Our starting point is the operator ansatz,
\begin{equation}
e^{-\Delta\tau U(\hat{n}_{\uparrow}-\frac{1}{2})(\hat{n}_{\downarrow}-\frac{1}{2})}=\int e^{a(s)\hat{\mathcal{O}}(s)}b(s)\,\mathd s,\label{eq:gen_hs1}
\end{equation}
with functions $a(s)$ and $b(s)$ as yet to be determined. The right-hand side introduces a real auxiliary field $s$ that couples to electron charge
or spin magnetic moment
\begin{equation}
\hat{\mathcal{O}}(s)=\begin{cases}
\hat{n}_{\uparrow}+\hat{n}_{\downarrow}-1 & (U<0)\\
\hat{n}_{\uparrow}-\hat{n}_{\downarrow} & (U>0)
\end{cases},\label{eq:gen_hs2}
\end{equation}
corresponding to attractive or repulsive Hubbard $U$, respectively.
The electron number operators $\hat{n}_{\uparrow}$ and $\hat{n}_{\downarrow}$ for a single site
have eigenvalues 0 or 1. The discretization in imaginary time $\Delta\tau=\beta/N_\tau$
is a tunable parameter, and originates from a Suzuki-Trotter expansion
of the partition function, whereby the inverse temperature $\beta$
is subdivided into $N_\tau$ parts~\cite{Blankenbecler81}.

Special cases of Eq.~\eqref{eq:gen_hs1} include the Gaussian transformation,
$a(s)\sim s$ and $b(s)\sim\exp(-s^{2})$, as originally considered
by Hubbard and Stratonovich~\cite{Hubbard59,Stratonovich58}, and the discrete transformation proposed by Hirsch, for which
the choice $b(s)\sim[\delta(s+1)+\delta(s-1)]$ effectively introduces
Ising auxiliary spins, $s=\pm1$~\cite{Hirsch83}. There is also extensive literature introducing other types of HS transformations and providing general rules for formulating new ones~\cite{Batrouni90, Chen92, Batrouni93, KarakuzuPRB2018}. 

DQMC simulations are frequently limited by the appearance of a sign problem~\cite{Loh90, Gubernatis16}. The severity of the sign problem depends partly on the choice of the HS transformation, with discrete auxiliary variables frequently being favorable. If the sign problem is not severe, an advantage of working with continuous HS fields is that they enable powerful sampling methods like Langevin dynamics and Hamiltonian Monte Carlo (HMC)~\cite{Duane87,Neal99}. 
These sampling methods can be helpful for reducing long autocorrelation times, especially near critical points. For example, continuous variables make possible the application of powerful
Fourier acceleration techniques, whereby the dynamical relaxation rate is adjusted according to imaginary-time freqeuency~\cite{Batrouni85}.
As another example, the fictitious momentum in
HMC yields inertial dynamics that
can reduce the dynamical critical exponent $z$~\cite{Kennedy01}.
Such sampling methods were originally developed in the context of
lattice gauge theory, and have more recently proved to be highly effective
for simulation of electron-phonon models~\cite{Beyl18, Batrouni19, Cohen-Stead22, CohenSteadPreprint}.

A continuous but compact HS transformation has been proposed
by D.~Lee~\cite{Lee08}; it blends some of the trade-offs of the
discrete and continuous HS transformations discussed above. In this
approach, auxiliary variables in the domain $s\in[-\pi,\pi]$ are coupled sinusoidally to the electrons.
This compact HS transformation was found to be the most efficient strategy for
simulating a dilute gas of attractive
fermions in the unitary limit. The auxiliary field, as continuous variables, could be sampled using the powerful HMC method. Furthermore, due to their compact range, this HS transformation yielded the best conditioning of the associated fermion matrices.

Here, we further develop this compact HS transformation approach in two ways. First,
we derive a general set of constraints on $a(s)$ and $b(s)$ such
that the HS transformation in Eq.~\eqref{eq:gen_hs1} is exact at
all orders in $\Delta\tau$, and use these constraints to derive systematic corrections to previous results. Second, following the suggestion of Lee, we introduce a class
of HS transformations that continuously interpolates from the compact, sinusoidal HS transformation to the discrete HS transformation of Hirsch. A final contribution of this paper is to benchmark the new class of HS transformations 
using DQMC simulations of the two-dimensional single-band square and triangular Hubbard models in the strong coupling limit. 

\section{A general class of HS transformations}

For notational convenience, we introduce the operator
\begin{equation}
\hat{m}_{\sigma}=2\hat{n}_{\sigma}-1,
\end{equation}
and represent the sign of $U$ as
\begin{equation}
\eta=U/|U|=\pm1.\label{eq:eta}
\end{equation}
The ansatz of Eqs.~\eqref{eq:gen_hs1} and~\eqref{eq:gen_hs2} may
now be written in the compact form
\begin{equation}
e^{-\frac{1}{4}\Delta\tau Um_{\uparrow}m_{\downarrow}}=\int e^{\frac{a(s)}{2}(\hat{m}_{\uparrow}-\eta\hat{m}_{\downarrow})}b(s)\,\mathd s.\label{eq:ansatz}
\end{equation}
Unless otherwise stated, the integration domain extends over the entire real line, $s \in \mathbb R$.

The operators $\hat{m}_{\uparrow}$ and $\hat{m}_{\downarrow}$ commute,
and each has two eigenvalues, $\pm1$. The local Hilbert space is spanned by the four eigenstates that simultaneously
diagonalize $m_{\uparrow}$ and $m_{\downarrow}$. Equation~\eqref{eq:ansatz} is valid if and only if the operators on both sides have an identical action when applied to each of these four
eigenstates. To achieve this, we may effectively replace the operators $(\hat{m}_\uparrow, \hat{m}_\downarrow)$ with all possible combinations of eigenvalues. The two cases, $(+1,-\eta)$ 
and $(-1, +\eta)$ yield the constraints
\begin{align}
e^{-\frac{1}{4}\Delta\tau|U|} & =\int e^{+a(s)}b(s)\,\mathd s,\label{eq:bare_const1}\\
e^{-\frac{1}{4}\Delta\tau|U|} & =\int e^{-a(s)}b(s)\,\mathd s.\label{eq:bare_const2}
\end{align}
The cases $(+1,+\eta)$ and $(-1,-\eta)$ yield an additional constraint
\begin{equation}
e^{\frac{1}{4}\Delta\tau|U|}=\int b(s)\,\mathd s.\label{eq:constraint1}
\end{equation}
Averaging Eqs.~\eqref{eq:bare_const1} and~\eqref{eq:bare_const2},
we find
\begin{equation}
e^{-\frac{1}{4}\Delta\tau|U|}=\int\cosh\left[a(s)\right]b(s)\,\mathd s.\label{eq:constraint2}
\end{equation}
Subtracting them yields
\begin{equation}
0=\int\sinh\left[a(s)\right]b(s)\,\mathd s.\label{eq:constraint3}
\end{equation}
Equations~\eqref{eq:constraint1}--\eqref{eq:constraint3} are necessary and sufficient conditions
for the correctness of the ansatz, Eq.~\eqref{eq:ansatz},
or equivalently, Eq.~\eqref{eq:gen_hs1}. Typically we will select $a(s)$ as an odd function, and $b(s)$ as an even function, such that Eq.~\eqref{eq:constraint3} is immediately satisfied.

Constraints analogous to Eqs.~\eqref{eq:constraint1}--\eqref{eq:constraint3} were previously derived in Appendix A2 of Ref.~\onlinecite{Wan2020}.

\section{Review of known HS transformations}
Let us now review how some existing HS transformations fit into the
form of Eq.~\eqref{eq:ansatz}.

\subsection{Gaussian auxiliary field\label{subsec:gaussian}}

For illustrative purposes, we will derive the Gaussian
HS transformation using a more standard procedure. The operator identity
\begin{equation}
\int e^{-\frac{1}{2}(s-\hat{A})^{2}}\,\mathd s=\int e^{-\frac{1}{2}s^{2}}\mathd s
\end{equation}
is valid for any Hermitian $\hat{A}$. To see this, one may work in the eigenbasis, such that $\hat{A}$ is effectively replaced by an arbitrary eigenvalue $\lambda$. The integral is invariant to the constant shift $s \rightarrow s + \lambda$, establishing the desired equality.

Expanding the square on the left, and performing the Gaussian integral
on the right, we find
\begin{equation}
e^{-\frac{1}{2}\hat{A}^{2}}\int e^{-\frac{1}{2}s^{2}+s\hat{A}}\,\mathd s=\sqrt{2\pi}.\label{eq:operator_integral}
\end{equation}
To make contact with Eq.~\eqref{eq:ansatz}, select
\begin{equation}
\hat{A}=\frac{1}{2}\sqrt{\Delta t|U|}\,(\hat{m}_{\uparrow}-\eta\hat{m}_{\downarrow}).
\end{equation}
The commutativity of $\hat{m}_\uparrow$ and $\hat{m}_\downarrow$, the identity $\hat{m}_\sigma^2 = 1$, and the identity $\eta |U| = U$ together yield, 
\begin{align}
\hat{A}^{2} & =\frac{\Delta t|U|}{2}-\frac{\Delta tU}{2}\hat{m}_{\uparrow}\hat{m}_{\downarrow}.
\end{align}
Inserting these results into Eq.~\eqref{eq:operator_integral} and
rearranging terms, we recover the ansatz of Eq.~\eqref{eq:ansatz},
where
\begin{align}
a(s) & =\sqrt{\Delta\tau|U|}\,s\\
b(s) & =\frac{1}{\sqrt{2\pi}}e^{-\frac{1}{2}s^{2}-\frac{1}{4}\Delta t|U|}.
\end{align}
One may verify that these functions satisfy
the integral constraints of Eqs.~\eqref{eq:constraint1} and~\eqref{eq:constraint2},
as expected.


\subsection{Ising auxiliary field\label{subsec:hirsch}}

Hirsch introduced the HS transformation~\cite{Hirsch83},
\begin{equation}
e^{-\frac{1}{4}\Delta\tau Um_{\uparrow}m_{\downarrow}}=\frac{1}{2}e^{-\frac{1}{4}\Delta\tau\left|U\right|}\sum_{s=\pm1}e^{\frac{\alpha s}{2}(m_{\uparrow}-\eta m_{\downarrow})},\label{eq:hirsch}
\end{equation}
where $s=\pm1$ is now an Ising auxiliary field. The real constant
$\alpha$ is defined to satisfy
\begin{equation}
\cosh\alpha=e^{\frac{1}{2}\Delta\tau\left|U\right|}.\label{eq:alpha_def}
\end{equation}
This takes the form of our ansatz, Eq.~\eqref{eq:ansatz}, upon defining
\begin{align}
a(s) & =\alpha s\label{eq:hirsch_a}\\
b(s) & =\frac{1}{2}e^{-\frac{1}{4}\Delta\tau\left|U\right|}\left[\delta(s+1)+\delta(s-1)\right].\label{eq:hirsch_b}
\end{align}
Again, one may verify that the constraints of Eqs.~\eqref{eq:constraint1}
and~\eqref{eq:constraint2} are satisfied.

\subsection{Compact auxiliary field with periodic coupling\label{subsec:lee}}

Lee proposed a compact HS transformation~\cite{Lee08}, which takes
the form of Eq.~\eqref{eq:ansatz} using the definitions

\begin{align}
a(s) & =\sqrt{c_0}\,\sin s\label{eq:lee_a}\\
b(s) & =\frac{1}{2\pi}e^{-\frac{1}{4}\Delta\tau\left|U\right|}\,\Theta(\pi-|s|).\label{eq:lee_b}
\end{align}
The Heaviside step function $\Theta(\cdot)$ constrains the integral
of Eq.~\eqref{eq:ansatz} to the compact domain $-\pi<s<\pi$. Using
the path integral formalism, Lee derived an approximate coefficient,
\begin{equation}
c_0\approx2\Delta\tau\left|U\right|.\label{eq:c_approx}
\end{equation}
Below, we will derive corrections to $c_0$ by expanding in powers of
the small parameter $\Delta\tau$. Such corrections are important
to maintain the overall $\mathcal{O}(\Delta\tau^{2})$ accuracy of 
a DQMC code.

Observe that the function $b(s)$ already satisfies the first constraint,
Eq.~\eqref{eq:constraint1}. The second constraint, Eq.~\eqref{eq:constraint2},
then determines $c_0$.

A general integral identity is
\begin{equation}
\frac{1}{2\pi}\int_{-\pi}^{\pi}\cosh\left(\sqrt{c_0}\sin s\right)\mathd s=I_{0}(\sqrt{c_0}),
\end{equation}
where $I_{\alpha}(x)=\mathrm{i}^{-\alpha}J_{\alpha}(\mathrm{i}\,x)$ is the modified Bessel
function of the first kind. This integral matches that appearing in Eq.~\eqref{eq:constraint2} given the definitions of $a(s)$ and $b(s)$. The resulting constraint is,
\begin{equation}
I_{0}(\sqrt{c_0})=e^{\frac{1}{2}\Delta\tau\left|U\right|}.\label{eq:discrete_constraint}
\end{equation}
Taylor expansion on the left and substitution of 
\begin{equation}
x=\Delta\tau\left|U\right|/2\label{eq:xdef}
\end{equation}
on the right yields an implicit equation for $c_0$,
\begin{equation}
\sum_{n=0}^{\infty}\frac{c_0^{n}}{n!^{2}4^{n}} = e^x.\label{eq:I0_exp}
\end{equation}
Note that $x$ can be made arbitrarily small through an appropriate choice of the discretization in imaginary time $\Delta \tau$. With the help of a symbolic algebra package, we find the series expansion,
\begin{equation}
c_0=4x+x^{2}+\frac{1}{18}x^{3}-\frac{1}{72}x^{4}+\frac{7}{10800}x^{5}+\dots.\label{eq:c_exp_compact}
\end{equation}


Observe that the first order approximation, $c_0\approx4x$, reproduces
Eq.~\eqref{eq:c_approx}. Truncation at this level is not advisable,
however, as the corresponding approximation to Eq.~\eqref{eq:lee_a}
becomes fairly imprecise,

\begin{equation}
a(s)=\sqrt{2\Delta\tau\left|U\right|}\sin s+\mathcal{O}(\Delta\tau^{3/2}).\label{eq:a_approx}
\end{equation}
This level of truncation error should be compared to the discretization
error already present in a DQMC simulation, which is globally of second order in $\Delta \tau$. This error originates from a Suzuki-Trotter
expansion involving symmetric operator splitting, $e^{\Delta \tau(\hat{A} + \hat{B})} \approx e^{\Delta \tau \hat{A}/2} e^{\Delta \tau \hat{B}} e^{\Delta \tau \hat{A}/2}$, which is locally
accurate to third order in $\Delta\tau$~\footnote{In describing DQMC codes, one frequently sees  written $e^{\tau (A+B)}\approx e^{\tau A} e^{\tau B}$, but typically this originates from symmetric operator splitting and application of the cyclic property of the trace.}. It appears, then, that retaining
more terms in the expansion of Eq.~\eqref{eq:c_exp_compact} is important
to the overall accuracy of a DQMC code.

\section{Interpolating between Ising and sinusoidal HS transformations \label{subsec:interp_hs}}

The constraints of Eqs.~\eqref{eq:constraint1} and~\eqref{eq:constraint2}
are relatively easy to satisfy, and allow great flexibility in designing
new HS transformations with the form of Eq.~\eqref{eq:ansatz}. For example, it is possible to continuously 
interpolate between the HS transformations of Secs.~\ref{subsec:hirsch}
and~\ref{subsec:lee} via
\begin{align}
a(s) & = \sqrt{c_p} \, \frac{\atan(p \sin s)}{\atan \, p} \label{eq:interp_a}\\
b(s) & =\frac{1}{2\pi}e^{-\frac{1}{4}\Delta\tau\left|U\right|}\,\Theta(\pi-|s|),\label{eq:interp_b}
\end{align}
where $0 < p < \infty$ is the interpolation parameter. The coefficient $c_p$ controls the coupling strength between the auxiliary field and fermions, and remains to be determined.

The limit $p \rightarrow 0$ recovers Eqs.~\eqref{eq:lee_a} and~\eqref{eq:lee_b}.
The limit $p \rightarrow \infty$ is a bit more subtle. The domain of $s$ may be restricted to $[-\pi, \pi]$, for which
\begin{equation}
\lim_{p\rightarrow \infty}a(s)=\sqrt{c_p}\,\frac{s}{|s|}.
\end{equation}
The integral anstaz
of Eq.~\eqref{eq:ansatz} becomes
\begin{equation}
\lim_{p\rightarrow\infty}\int e^{\frac{a(s)}{2}(\hat{m}_{\uparrow}-\eta\hat{m}_{\downarrow})}b(s)\,\mathd s=\frac{1}{2}e^{-\frac{\Delta\tau\left|U\right|}{4}}(I^{+}+I^{-}),
\end{equation}
where
\begin{equation}
I^{\pm}=e^{\pm\frac{1}{2}\sqrt{c_p}(\hat{m}_{\uparrow}-\eta\hat{m}_{\downarrow})}\left(\frac{1}{\pi}\int_{\Omega^{\pm}}\,\mathd s\right),\label{eq:I_pm}
\end{equation}
and $s$ is to be sampled from the two sub-domains, $\Omega^{-}=[-\pi,0]$ and $\Omega^{+}=[0,\pi]$. In the
context of a DQMC code, the sampling weight depends only on whether
$s\in\Omega^{-}$ or $s\in\Omega^{+}$. That is, we could effectively
replace each of these two continuous sampling domains with just two
allowed values, $s=\pm1$. As pointed out in Ref.~\onlinecite{Lee08}, this limit recovers the discrete Ising HS transformation,
Eqs.~\eqref{eq:lee_a} and~\eqref{eq:lee_b}, where $\sqrt{c_\infty}=\alpha$,
as defined in Eq.~\eqref{eq:alpha_def}.

\begin{figure}
\includegraphics[width=0.95\columnwidth]{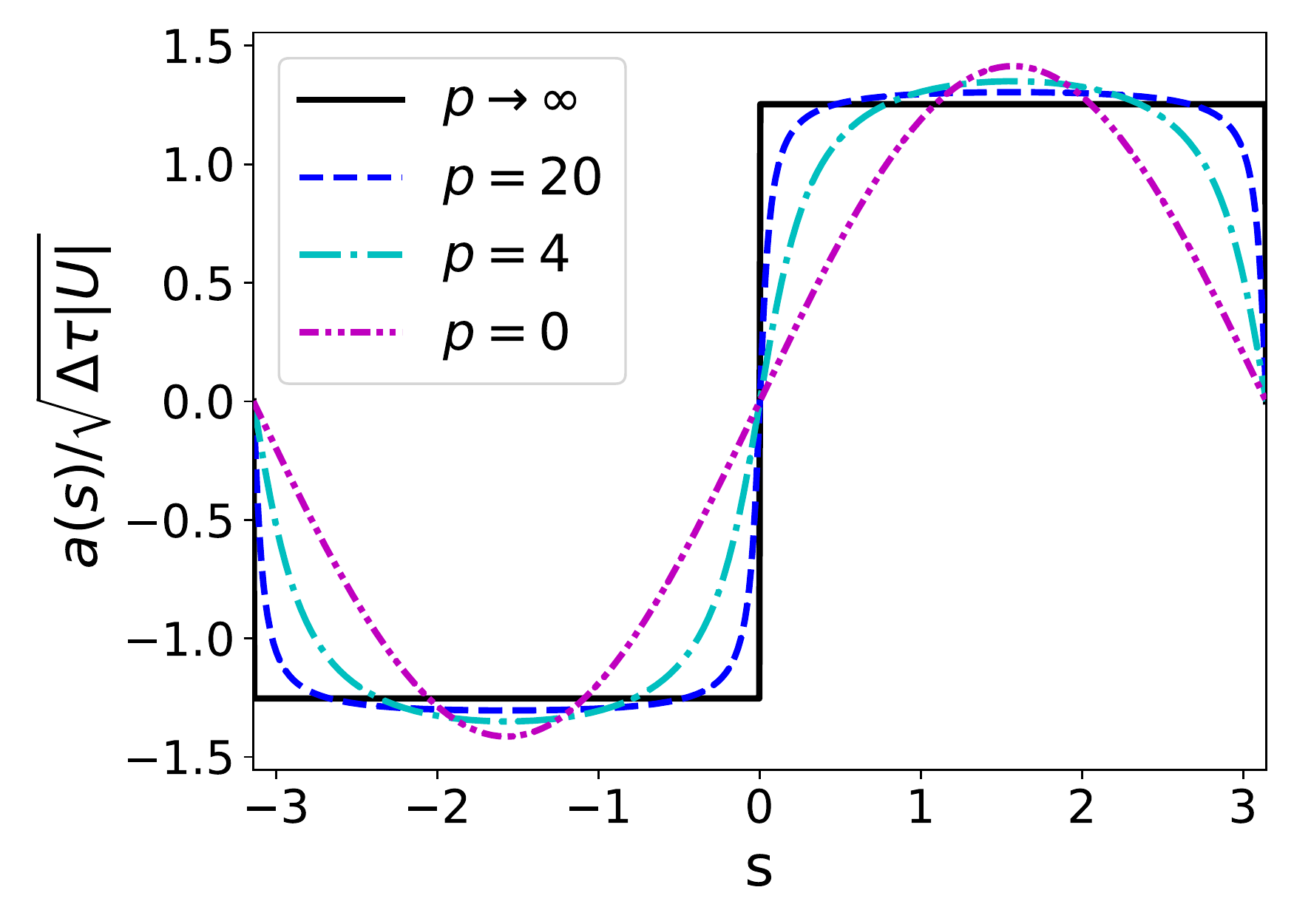}
\caption{A class of compact Hubbard Stratonovich transformations defined by Eqs.~\eqref{eq:interp_a} and~\eqref{eq:interp_b}. Each curve $a(s)$ defines a possible coupling between electrons and the auxiliary field $s$.
The square wave coupling ($p\rightarrow \infty$) effectively corresponds to the usual
Ising auxiliary variables $s =\pm 1$~\cite{Hirsch83}. The sine wave coupling ($p \rightarrow 0$)
is the periodic HS transformation introduced by D.~Lee~\cite{Lee08}.\label{fig:interpolate}}
\end{figure}

Figure~\ref{fig:interpolate} illustrates how varying $p$ from
0 to $\infty$ interpolates between previously known HS transformations. Here we selected $c_p$ according to its $\Delta \tau \rightarrow 0$ limit [Eq.~\eqref{eq:c0}], which will be derived below.

The constant $c_p$ is, in general, determined by the constraint
Eq.~\eqref{eq:constraint2}, which may be written
\begin{equation}
\frac{1}{2\pi}\int_{-\pi}^{+\pi}\cosh\left(\sqrt{c_p} \, \frac{\atan(p \sin s)}{\atan \, p}\right)\,\mathd s=e^x,\label{eq:const-1}
\end{equation}
with $x$ defined in Eq.~\eqref{eq:xdef}. In applications, we will typically have numerical values for $p$ and $x$, and it is straightforward to solve for $c_p$ numerically; a Julia routine is provided in Appendix~\ref{sec:julia}.

One can formally Taylor expand $c_p(x)$ in small $x$, generalizing Eq.~\eqref{eq:c_exp_compact} to nonzero $p$. We will work out the leading order approximation. Using
\begin{equation}
\cosh a=\sum_{n=0}^{\infty}\frac{a^{2n}}{(2n)!},
\end{equation}
and $x= \Delta \tau |U| / 2$, expand both sides of Eq.~\eqref{eq:const-1} in small $c_p \sim \Delta \tau$,
\begin{equation}
    \frac{c_p}{4\pi}\int_{-\pi}^{+\pi}\left(\frac{\atan(p \sin s)}{\atan \, p}\right)^2\,\mathd s = \frac{\Delta \tau |U|}{2} + \mathcal{O}(\Delta \tau^2).
\end{equation}
The limiting behavior for small $\Delta \tau$ is
\begin{equation}
\lim_{\Delta \tau \rightarrow0} \, \frac{c_p}{\Delta \tau |U|} = \left[ \frac{1}{2\pi}\int_{-\pi}^{+\pi}\left(\frac{\atan(p \sin s)}{\atan \, p}\right)^2\,\mathd s \right]^{-1}.\label{eq:c0}
\end{equation}
The right-hand side decreases monotonically as a function of the interpolation parameter $p$. For example,
\begin{equation}
        \lim_{\Delta \tau \rightarrow0} \, \frac{c_p}{\Delta \tau |U|} = \begin{cases}
2 & (p = 0)\\
1.37546\dots & (p = 4) \\
1.11849\dots & (p = 20) \\
1 & (p = \infty)
\end{cases},
\end{equation}
where the first and last cases should be understood as limits. The $p = 0$ result is consistent with Eq.~\eqref{eq:c_exp_compact}, and the $p = \infty$ result with Eq.~\eqref{eq:alpha_def} where $\alpha = \sqrt{c_\infty}$. Observe that increasing $p$, i.e. moving toward the discrete Ising HS transformation, effectively decreases the coupling strength $\sqrt{c_p}$ between the auxiliary field and the fermions.

\section{Numerical benchmarks}

We explore performance of the proposed compact HS transformations in the context of the doped single-band Hubbard Hamiltonian
\begin{equation}\label{eq:H}
    H = -t\sum_{\langle i,j \rangle,\sigma} 
    c^\dagger_{i,\sigma}c^{\phantom\dagger}_{j,\sigma}
    - \mu \sum_{i,\sigma} n_{i,\sigma} + 
    U \sum_i n_{i,\uparrow} n_{i,\downarrow}. 
\end{equation}
Here, $c^\dagger_{i,\sigma}$ ($c^{\phantom\dagger}_{i,\sigma}$) is the creation (annihilation) operator for a spin-$\sigma$ ($=\uparrow,\downarrow$) electron on lattice site $i$, $t$ is the hopping integral for nearest neighbor sites $\langle i,j \rangle$, $\mu$ is the chemical potential, and $U$ is the onsite Hubbard repulsion. We consider the model defined on two-dimensional square and triangular lattices with $N = L\times L$ sites and arbitrary in-plane lattice constants. 
For all simulations, the discretization in imaginary time is $\Delta \tau = 0.1/t$. 
The Monte Carlo sampling task is to generate auxiliary fields  $s_{i,\tau}$ according to the weight $\exp(-S) = | \det M_\uparrow \det M_\downarrow |$. Each  $M_\sigma$ is an $N \times N$ matrix function of the fields $s_{i,\tau}$, and this functional dependence varies according to the choice of HS transformation~\cite{Blankenbecler81, WhitePRB1989, Gubernatis16}.

The compact HS transformations of Sec.~\ref{subsec:interp_hs} are tunable by a parameter $p$. The limit $p \rightarrow \infty$ gives rise to the usual discrete Ising HS transformation (Sec.~\ref{subsec:hirsch}). Simulations in this limit are efficiently performed using the traditional DQMC approach ~\cite{Blankenbecler81, WhitePRB1989}. The method sweeps over all imaginary time-slices, and within each, all lattice sites. At each space-time point $(i,\tau)$, a single spin-flip $s_{i,\tau} \rightarrow -s_{i,\tau}$ is proposed and then accepted with Metropolis probability $\min[1,\exp(-\Delta S)]$, where $\Delta S$ denotes the associated change in action. After a successful spin-flip, local data structures [the equal-time Green's functions $G(\tau)$] are updated at an amortized cost that scales approximately like $\mathcal{O}(N^2)$. The cost to fully sweep over all $N_\tau$ imaginary times and $N$ sites then scales like $\mathcal{O}(N^3 N_\tau)$. Numerical errors can accumulate when sweeping through time slices, and one must periodically recompute the equal-time Green's function using a numerical stabilization procedure~\cite{WhitePRB1989, Loh1992, Loh2005, Gubernatis16, Bauer2020}. 

\begin{figure}
    \centering
    \includegraphics[width=\columnwidth]{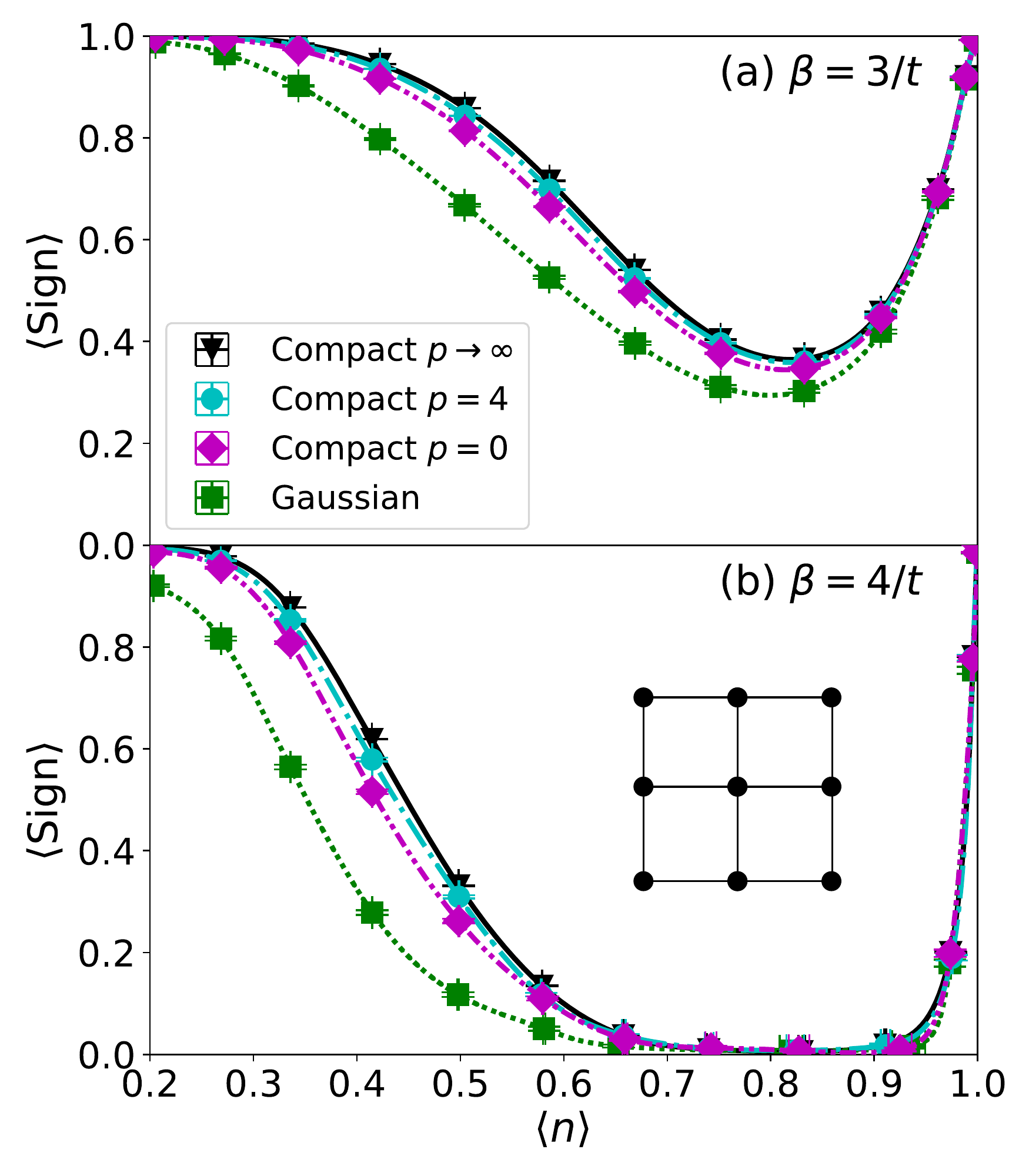}
    \caption{The average sign of $\det M_\uparrow\det M_\downarrow$ as a function of electron filling  $\langle n \rangle$ for the two-dimensional single band Hubbard model on a square lattice with $U = 8t$, $N = 8\times 8$, and (a) $\beta = 3/t$ or (b)  $\beta=4/t$. Various HS transformations are compared, with the Ising limit ($p \rightarrow \infty$) generally producing the best sign. }
    \label{fig:fermion_sign}
\end{figure}

Alternatively, when $p$ is finite, each auxiliary variable $s_{i,\tau}$ can be viewed as a continuous degree of freedom in the periodic domain $[-\pi, \pi]$. The coupling strength $c_p$ between auxiliary field and electrons is determined by Eq.~\eqref{eq:c0}, which can be solved using the Julia code in Appendix~\ref{sec:julia}. In our numerical implementation, we opted to sample the field $s_{i,\tau}$ using the Langevin Monte Carlo method~\cite{Kennedy90} (or equivalently, the Metropolis-adjusted Langevin method~\cite{Besag1994}). This approach can be understood as a variant of HMC where each dynamical trajectory consists of only a single-time step~\cite{Neal99}. The method associates with each field component $s_{i,\tau}$ a fictitious velocity $v_{i,\tau}$. A trial update of the auxiliary field $s_{i,\tau} \rightarrow s^\prime_{i,\tau}$ has two steps. First, one samples all velocities $v_{i,\tau}$ from the Gaussian equilibrium distribution. Second, one performs Verlet integration for a single time-step $\epsilon$,
\begin{align}
    s^\prime &= s - \frac{\epsilon^2}{2} \nabla S + \epsilon v \\
    v^\prime &= v - \frac{\epsilon}{2} \nabla S - \frac{\epsilon}{2} \nabla S^\prime.
\end{align}
Here, $S$ and $S^\prime$ denote the action evaluated at fields $s$ and $s^\prime$, respectively, and $\nabla$ denotes the gradient with respect to $s_{i,\tau}$ at every space-time index. The detailed balance condition is achieved by accepting the proposed update with probability
\begin{equation}
    P_\mathrm{accept} = \min [1,\exp (-\Delta S - \Delta K)].
\end{equation}
As before, $\Delta S = S^\prime - S$ represents the change in action. Additionally, we must include a term $\Delta K = \frac{1}{2} \sum_{i,\tau} (v^{\prime 2}_{i,\tau} - v^2_{i,\tau})$ representing the change in fictitious kinetic energy. The dominant numerical cost in each Langevin step is the calculation of the new action $S^\prime$ and its derivative $\nabla S^\prime$. As with DQMC, numerical stabilization is necessary, and the computational cost for a full system update again scales like $\mathcal O(N^3 N_\tau)$.

Figure~\ref{fig:fermion_sign} presents the results from simulations  
of the square lattice Hubbard model for
various HS transformations as a function of the parameter $p$ and $U = 8t$. The $x$-axis shows the estimated mean electron number $\langle n \rangle$, which is indirectly controlled by a varying chemical potential $\mu$. For example, the data at half-filling, $\langle n \rangle =1$, was generated using $\mu = 0$, and the data at $\langle n \rangle \approx 0.668$ was generated using $\mu = -3.5t$. The $y$-axis shows the expected value of
\begin{equation}
    \mathrm{Sign} = \frac{\det M_\uparrow \det M_\downarrow}{|\det M_\uparrow \det M_\downarrow|}.
\end{equation}
The deviation of $\langle \mathrm{Sign} \rangle$ from one is a proxy for the difficulty of the so-called sign problem~\cite{Loh90, Troyer05, IglovikovPRB2015, Iazzi16, mondaini22}. All HS transformations yield the same qualitative behavior, which is consistent with previous results obtained for the square lattice Hubbard model using Ising auxiliary fields~\cite{WhitePRB1989, IglovikovPRB2015}.  Particle-hole symmetry at half-filling perfectly protects against the sign problem. Upon reducing $\langle n \rangle$ from one, the average sign rapidly decreases until it hits a broad minimum value around $\langle n \rangle \approx 0.8$. As the filling is reduced further, the average sign begins to slowly recover before reaching one in the dilute limit. 

The discrete Hirsch HS transformation is reached in the limit where $p \rightarrow \infty$, and restricts each auxiliary variable $s_{i,\tau}$ to two possible values, $\pm 1$. This limit is observed to be the best at mitigating the sign problem (i.e. it produces the largest average sign at all fillings). Conversely, the continuous Gaussian transformation gives rise to the worst sign problem. The sinusoidal coupling studied by Lee ($p = 0$) achieves an average sign that is already quite close to the discrete case ($p = \infty$). By increasing the interpolation parameter $p$, it is possible to approach the discrete HS transformation arbitrarily closely, while retaining the continuous nature of the HS field $s \in [-\pi, \pi]$. 

\begin{figure}
    \centering
    \includegraphics[width=0.8\columnwidth]{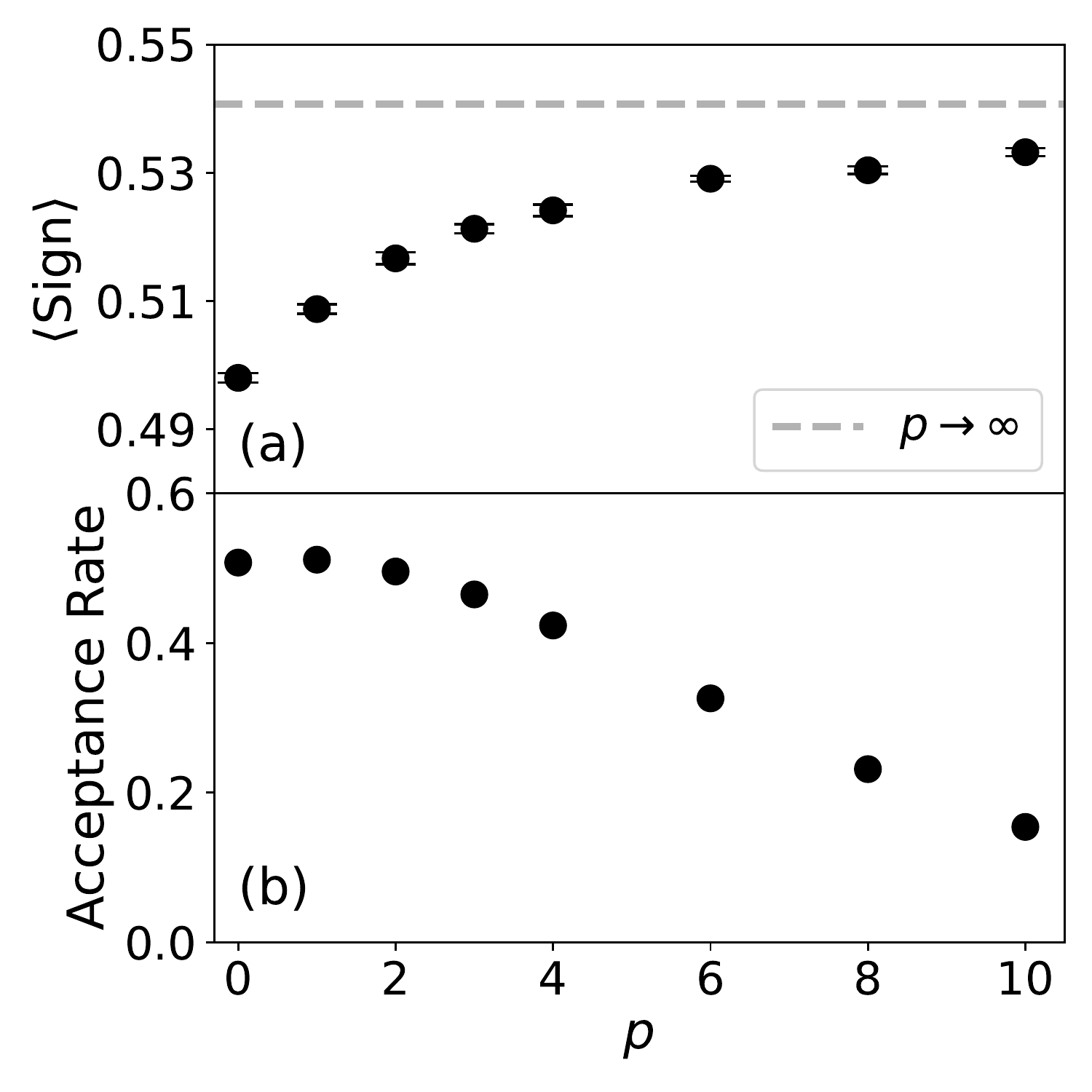}
    \caption{(a) The average sign of $\det M_\uparrow \det M_\downarrow$ and (b) the corresponding acceptance rate for proposed Langevin updates, for varying HS transformations as controlled by the parameter $p$. As in Fig.~\ref{fig:fermion_sign}a, we consider a square lattice Hubbard model with $U=8t,$ $\beta = 3/t$ and $N = 8 \times 8$, but here we focus on $\langle n \rangle \approx 0.668$, corresponding to $\mu = -3.5t$.}
    \label{fig:varied_p}
\end{figure}

Figure~\ref{fig:varied_p}a shows a closer view of the $p$-dependence on the average sign. Here, we fix $\mu = -3.5t$, corresponding to a particularly difficult filling fraction of $\langle n \rangle \approx 0.668$. (All other simulation parameters are identical to those used in Fig.~\ref{fig:fermion_sign}.) The average sign i ncreases monotonically with $p$.  

Langevin or HMC sampling methods decorrelate fastest when $p$ is of order one. Because forces are proportional to $\mathrm{d}a/\mathrm{d}s$, they are either vanishing or divergent when $p \rightarrow \infty$. Figure~\ref{fig:varied_p}b shows that, for a fixed integration time-step of  $\epsilon = 0.1$, the acceptance rate for proposed Langevin updates steadily decreases with increasing $p$. Furthermore, at large $p$, each accepted update become less effective in decorrelating the auxiliary field, because the typical forces are very small. In numerical practice, the optimal choice of $p$ should balance the benefits of increasing the average sign against the disadvantages of reducing autocorrelation time in the dynamical sampling scheme.

\begin{figure}[t]
    \centering
    \includegraphics[width=\columnwidth]{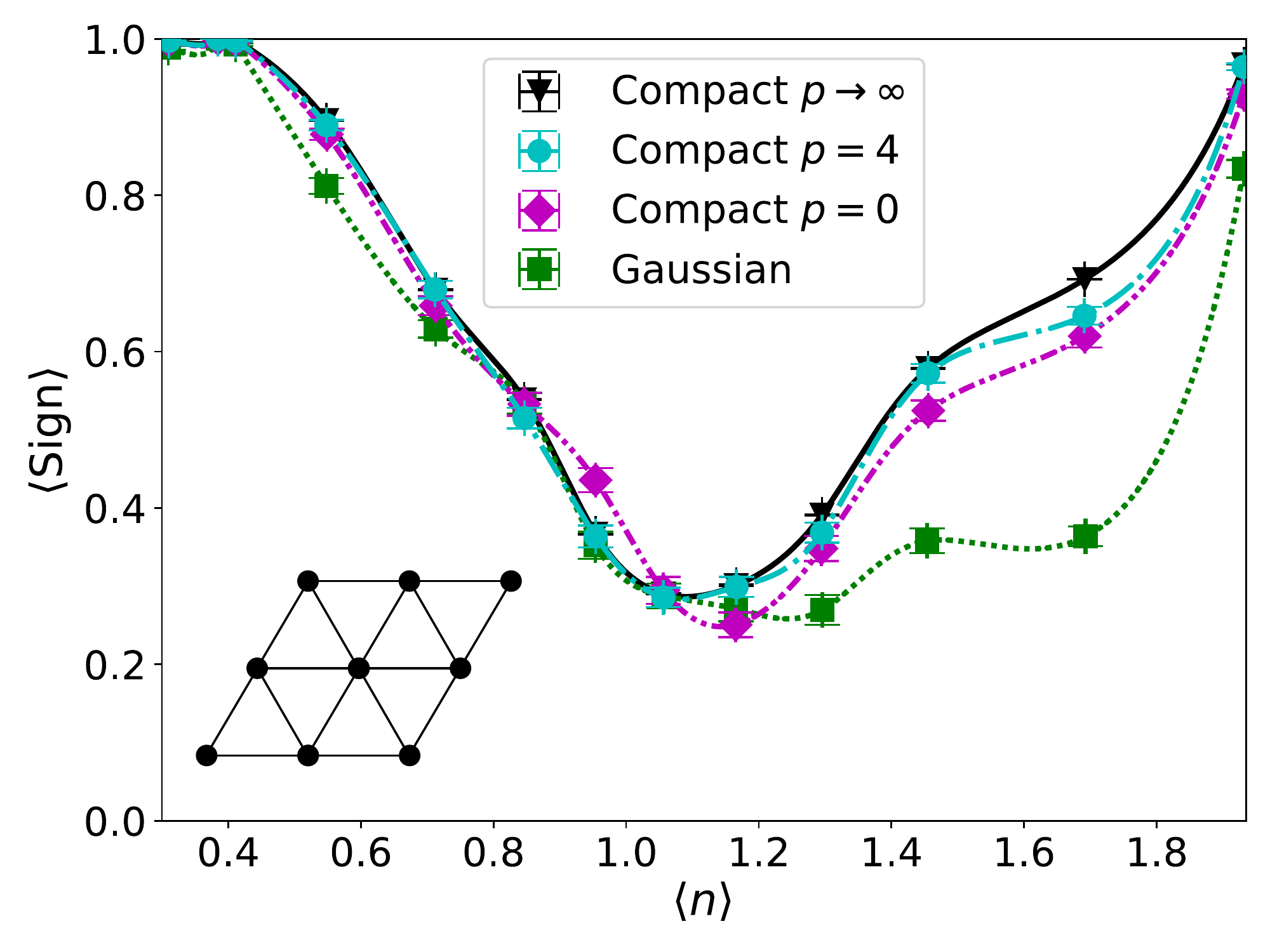}
    \caption{The average sign of $\det M_\uparrow\det M_\downarrow$ as a function of electron filling  $\langle n \rangle$ for the triangular lattice Hubbard model with $U = 6t$, $N = 6\times 6$ and $\beta = 3.5/t$. Various HS transformations are compared, with discrete Ising variables emerging when $p\rightarrow \infty$. Dotted curves show cubic spline interpolation as a guide to the eye. }
    \label{fig:triangular}
\end{figure}

Figure~\ref{fig:triangular} shows results obtained for the triangular lattice single band Hubbard model. Since this generally exhibits a more severe sign problem compared to the square lattice case~\cite{IglovikovPRB2015}, we focus here on an $N = 6\times 6$ lattice with $U = 6t$ and $\beta = 3.5/t$. The overall trends are very similar to those already discussed for the square lattice. The Guassian field produces the lowest average sign values at all carrier concentrations. The compact fields, on the other hand, produce larger average sign values at all carrier concentrations, and systematically approach the values obtained using Ising HS fields as $p$ increases. Interestingly, we also observe a small region $0.8 \le \langle n \rangle \le 1$ where the  $p=0$ compact field performs slightly better than the Ising fields.

In these simulations of the single band repulsive Hubbard models, the DQMC method required only a few sweeps to generate a decorrelated sample of the auxiliary field. In contrast, Langevin required two orders of magnitude more full-system updates to achieve comparable decorrelation. In the presence of a sign problem, the Langevin approach is at a fundamental disadvantage: The sampling weights $|\det M_\uparrow \det M_\downarrow|$ vanish upon each reversal of the sign in Eq.~\eqref{fig:fermion_sign}. This nodal surface corresponds to a logarithmically divergent action $S$ which, in principle, should disallow crossing by any continuous trajectory. In practice, the finite Langevin integration timestep $\epsilon$ makes crossing possible but rare. Previous work explored complexification of the auxiliary to enable continuous paths {\it around} the nodal surface, thereby avoiding ergodicity issues~\cite{Beyl18}. In future studies, it would be interesting to explore whether such complexification might be used in conjunction with the compact HS transformations of Eqs.~\eqref{eq:interp_a} and~\eqref{eq:interp_b}.

\section{Conclusions}

This work studies a class of HS transformations that continuously interpolates between the discrete Ising auxiliary variables introduced by Hirsch ($p = \infty$)~\cite{Hirsch83}, and the compact variables with periodic coupling introduced by Lee ($p = 0$)~\cite{Lee08}. As a proof of principle, we compared DQMC simulations ($p = \infty$) with Langevin simulations at various $p$, and found that the average sign systematically increases with $p$. Note, however, that these measurements are context dependent; the average sign is known to depend sensitively on the system's dimensionality \cite{IglovikovPRB2015, LiPRB2018}, its orbital basis \cite{KungPRB2016, LiPRB2018, Mou2022}, and the presence of additional interactions \cite{JohnstonPRB2013, KarakuzuPRB2018}. 

Although the sign problem is best mitigated at infinite $p$, selecting instead moderate $p$ enables the use of continuous sampling methods such as Langevin or HMC. In our study of the single band Hubbard model, we did not find benefit from Langevin sampling; this is partly because DQMC is already so effective at generating decorrelated samples, and partly because continuous sampling methods do not do well in crossing nodal surfaces. In other contexts, however, continuous sampling methods are known to significantly reduce long decorrelation times~\cite{Batrouni85,Kennedy01}. Previous studies of the attractive Hubbard model in the dilute limit found significant advantages to using HMC in conjunction with the compact $p=0$ HS transformation~\cite{Lee08}, and future work may benefit by selecting $p > 0$.  The use of continuous sampling methods also presents the possibility of using sparse iterative solvers to achieve near linear-scaling of computational cost with system size~\cite{Bai09,Beyl18,Cohen-Stead22}. Although linear-scaling simulations of the Hubbard model in the strongly-correlated limit is still not practical, the present study represents progress towards this direction.

Another context where a continuous HS transformation for the Hubbard interaction may be beneficial is in simulations of correlated systems with strong electron-phonon interactions. Langevin and HMC are known to be highly effective in sampling decorrelated phonon fields~\cite{Beyl18,Batrouni19,Cohen-Stead22}. Future work could perform {\em simultaneous} dynamical sampling of the phonon and HS auxiliary fields. Such an approach could prove useful in situations where the electron and phonon degrees of freedom are strongly coupled, e.g. small polarons.

\acknowledgments

This work was supported by the U.S. Department of Energy, Office of Science, Office of Basic Energy Sciences, under Award Number DE-SC0022311. This research used resources of the Oak Ridge Leadership Computing Facility, which is a DOE Office of Science User Facility supported under Contract No. DE-AC05-00OR22725.

\appendix

\begin{widetext}
\section{Numerical calculation of coupling strength} \label{sec:julia}

The Julia code below will solve Eq.~\eqref{eq:const-1} for the unknown $c_p$, taking as inputs $x = \Delta \tau |U| /2$ and finite interpolation parameter $p$.

{\large
\jlinputlisting{calculate_c.jl}
}

\end{widetext}

Numerical issues will arise at small $p$ due to the removable singularity at $p=0$,
\begin{equation*}
    \lim_{p\rightarrow 0} \atan(p \sin s)/\atan(p) = \sin s.
\end{equation*}
One solution is to expand the integrand powers of small $p$. Alternatively, when $p=0$ exactly, the coefficient $c_0$ can be calculated via the small-$x$ expansion of Eq.~\eqref{eq:c_exp_compact}.

\section{Additional results for the average sign}
\begin{figure}
    \centering
    \includegraphics[width=\columnwidth]{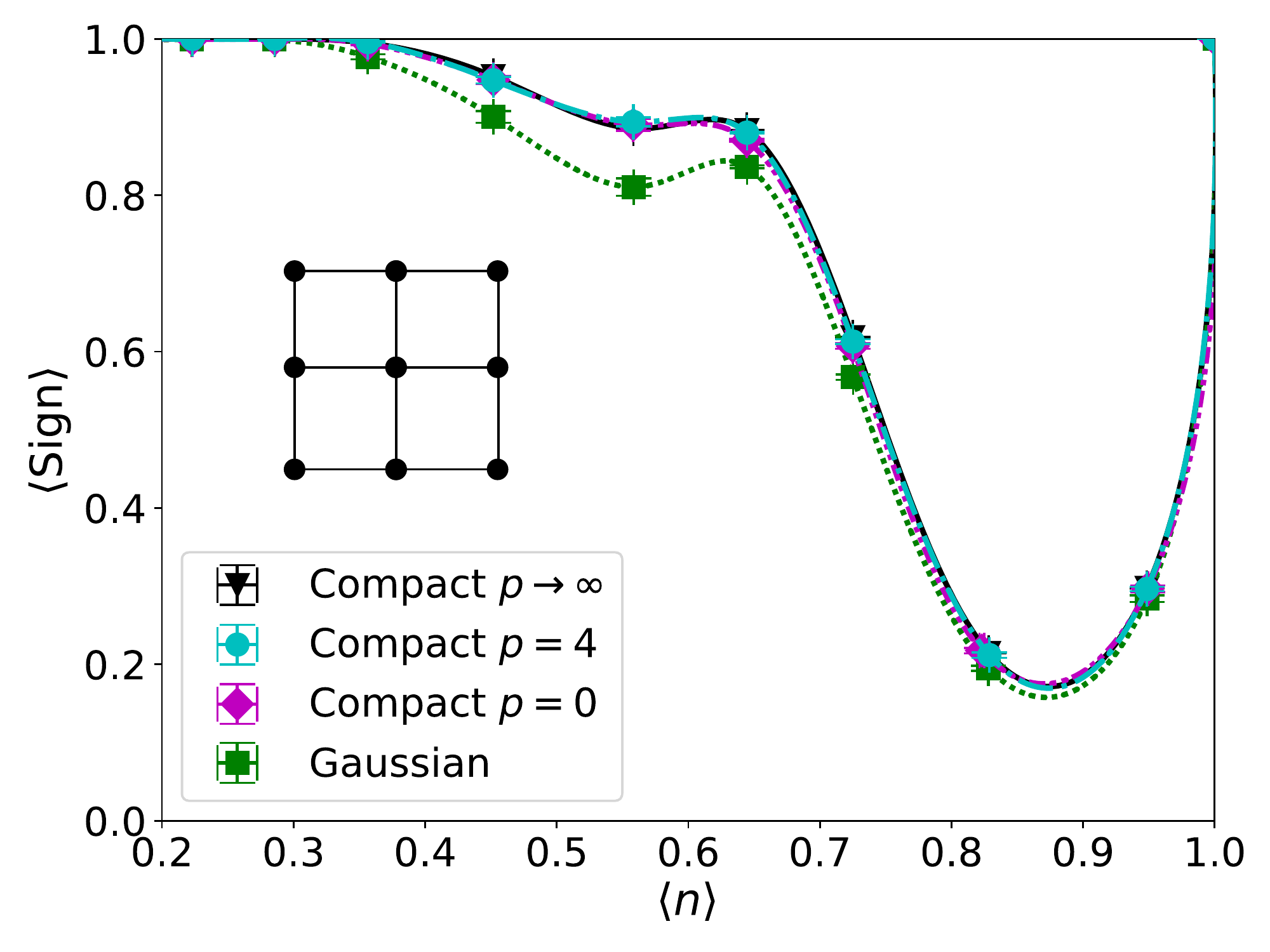}
    \caption{The average sign of $\det M_\uparrow\det M_\downarrow$ as a function of electron filling  $\langle n \rangle$ for the two-dimensional single band Hubbard model on a square lattice with $U = 4t$, $N = 8\times 8$ and $\beta = 4/t$.}
    \label{fig:square_U4_b6_L8}
\end{figure}

Figure~\ref{fig:square_U4_b6_L8} shows additional results for the square lattice Hubbard model on an $N = 8\times 8$ lattice with $U=4t$ and $\beta = 4/t$. The results resemble those presented in Fig.~\ref{fig:fermion_sign} in that the average sign for the Gaussian HS transformation has the smallest value across the full range of sampled densities.

\bibliography{references}

\begin{thebibliography}{36}%
\makeatletter
\providecommand \@ifxundefined [1]{%
 \@ifx{#1\undefined}
}%
\providecommand \@ifnum [1]{%
 \ifnum #1\expandafter \@firstoftwo
 \else \expandafter \@secondoftwo
 \fi
}%
\providecommand \@ifx [1]{%
 \ifx #1\expandafter \@firstoftwo
 \else \expandafter \@secondoftwo
 \fi
}%
\providecommand \natexlab [1]{#1}%
\providecommand \enquote  [1]{``#1''}%
\providecommand \bibnamefont  [1]{#1}%
\providecommand \bibfnamefont [1]{#1}%
\providecommand \citenamefont [1]{#1}%
\providecommand \href@noop [0]{\@secondoftwo}%
\providecommand \href [0]{\begingroup \@sanitize@url \@href}%
\providecommand \@href[1]{\@@startlink{#1}\@@href}%
\providecommand \@@href[1]{\endgroup#1\@@endlink}%
\providecommand \@sanitize@url [0]{\catcode `\\12\catcode `\$12\catcode
  `\&12\catcode `\#12\catcode `\^12\catcode `\_12\catcode `\%12\relax}%
\providecommand \@@startlink[1]{}%
\providecommand \@@endlink[0]{}%
\providecommand \url  [0]{\begingroup\@sanitize@url \@url }%
\providecommand \@url [1]{\endgroup\@href {#1}{\urlprefix }}%
\providecommand \urlprefix  [0]{URL }%
\providecommand \Eprint [0]{\href }%
\providecommand \doibase [0]{https://doi.org/}%
\providecommand \selectlanguage [0]{\@gobble}%
\providecommand \bibinfo  [0]{\@secondoftwo}%
\providecommand \bibfield  [0]{\@secondoftwo}%
\providecommand \translation [1]{[#1]}%
\providecommand \BibitemOpen [0]{}%
\providecommand \bibitemStop [0]{}%
\providecommand \bibitemNoStop [0]{.\EOS\space}%
\providecommand \EOS [0]{\spacefactor3000\relax}%
\providecommand \BibitemShut  [1]{\csname bibitem#1\endcsname}%
\let\auto@bib@innerbib\@empty
\bibitem [{\citenamefont {Blankenbecler}\ \emph {et~al.}(1981)\citenamefont
  {Blankenbecler}, \citenamefont {Scalapino},\ and\ \citenamefont
  {Sugar}}]{Blankenbecler81}%
  \BibitemOpen
  \bibfield  {author} {\bibinfo {author} {\bibfnamefont {R.}~\bibnamefont
  {Blankenbecler}}, \bibinfo {author} {\bibfnamefont {D.~J.}\ \bibnamefont
  {Scalapino}},\ and\ \bibinfo {author} {\bibfnamefont {R.~L.}\ \bibnamefont
  {Sugar}},\ }\bibfield  {title} {\bibinfo {title} {Monte {C}arlo calculations
  of coupled boson-fermion systems. {I}},\ }\href
  {https://doi.org/10.1103/PhysRevD.24.2278} {\bibfield  {journal} {\bibinfo
  {journal} {Phys. Rev. D}\ }\textbf {\bibinfo {volume} {24}},\ \bibinfo
  {pages} {2278} (\bibinfo {year} {1981})}\BibitemShut {NoStop}%
\bibitem [{\citenamefont {White}\ \emph {et~al.}(1989)\citenamefont {White},
  \citenamefont {Scalapino}, \citenamefont {Sugar}, \citenamefont {Loh},
  \citenamefont {Gubernatis},\ and\ \citenamefont {Scalettar}}]{WhitePRB1989}%
  \BibitemOpen
  \bibfield  {author} {\bibinfo {author} {\bibfnamefont {S.~R.}\ \bibnamefont
  {White}}, \bibinfo {author} {\bibfnamefont {D.~J.}\ \bibnamefont
  {Scalapino}}, \bibinfo {author} {\bibfnamefont {R.~L.}\ \bibnamefont
  {Sugar}}, \bibinfo {author} {\bibfnamefont {E.~Y.}\ \bibnamefont {Loh}},
  \bibinfo {author} {\bibfnamefont {J.~E.}\ \bibnamefont {Gubernatis}},\ and\
  \bibinfo {author} {\bibfnamefont {R.~T.}\ \bibnamefont {Scalettar}},\
  }\bibfield  {title} {\bibinfo {title} {Numerical study of the two-dimensional
  {Hubbard} model},\ }\href {https://doi.org/10.1103/PhysRevB.40.506}
  {\bibfield  {journal} {\bibinfo  {journal} {Phys. Rev. B}\ }\textbf {\bibinfo
  {volume} {40}},\ \bibinfo {pages} {506} (\bibinfo {year} {1989})}\BibitemShut
  {NoStop}%
\bibitem [{\citenamefont {Gubernatis}\ \emph {et~al.}(2016)\citenamefont
  {Gubernatis}, \citenamefont {Kawashima},\ and\ \citenamefont
  {Werner}}]{Gubernatis16}%
  \BibitemOpen
  \bibfield  {author} {\bibinfo {author} {\bibfnamefont {J.}~\bibnamefont
  {Gubernatis}}, \bibinfo {author} {\bibfnamefont {N.}~\bibnamefont
  {Kawashima}},\ and\ \bibinfo {author} {\bibfnamefont {P.}~\bibnamefont
  {Werner}},\ }\href@noop {} {\emph {\bibinfo {title} {Quantum Monte Carlo
  Methods}}}\ (\bibinfo  {publisher} {Cambridge University Press},\ \bibinfo
  {year} {2016})\BibitemShut {NoStop}%
\bibitem [{\citenamefont {Hubbard}(1959)}]{Hubbard59}%
  \BibitemOpen
  \bibfield  {author} {\bibinfo {author} {\bibfnamefont {J.}~\bibnamefont
  {Hubbard}},\ }\bibfield  {title} {\bibinfo {title} {Calculation of
  {{Partition Functions}}},\ }\href {https://doi.org/10.1103/PhysRevLett.3.77}
  {\bibfield  {journal} {\bibinfo  {journal} {Phys. Rev. Lett.}\ }\textbf
  {\bibinfo {volume} {3}},\ \bibinfo {pages} {77} (\bibinfo {year}
  {1959})}\BibitemShut {NoStop}%
\bibitem [{\citenamefont {Stratonovich}(1958)}]{Stratonovich58}%
  \BibitemOpen
  \bibfield  {author} {\bibinfo {author} {\bibfnamefont {R.}~\bibnamefont
  {Stratonovich}},\ }\bibfield  {title} {\bibinfo {title} {On a method of
  calculating quantum distribution functions},\ }\href@noop {} {\bibfield
  {journal} {\bibinfo  {journal} {Soviet Phys. Doklady}\ }\textbf {\bibinfo
  {volume} {2}},\ \bibinfo {pages} {416} (\bibinfo {year} {1958})}\BibitemShut
  {NoStop}%
\bibitem [{\citenamefont {Hirsch}(1983)}]{Hirsch83}%
  \BibitemOpen
  \bibfield  {author} {\bibinfo {author} {\bibfnamefont {J.~E.}\ \bibnamefont
  {Hirsch}},\ }\bibfield  {title} {\bibinfo {title} {Discrete
  {{Hubbard-Stratonovich}} transformation for fermion lattice models},\ }\href
  {https://doi.org/10.1103/PhysRevB.28.4059} {\bibfield  {journal} {\bibinfo
  {journal} {Phys. Rev. B}\ }\textbf {\bibinfo {volume} {28}},\ \bibinfo
  {pages} {4059} (\bibinfo {year} {1983})}\BibitemShut {NoStop}%
\bibitem [{\citenamefont {Batrouni}\ and\ \citenamefont
  {Scalettar}(1990)}]{Batrouni90}%
  \BibitemOpen
  \bibfield  {author} {\bibinfo {author} {\bibfnamefont {G.~G.}\ \bibnamefont
  {Batrouni}}\ and\ \bibinfo {author} {\bibfnamefont {R.~T.}\ \bibnamefont
  {Scalettar}},\ }\bibfield  {title} {\bibinfo {title} {Anomalous decouplings
  and the fermion sign problem},\ }\href
  {https://doi.org/10.1103/PhysRevB.42.2282} {\bibfield  {journal} {\bibinfo
  {journal} {Phys. Rev. B}\ }\textbf {\bibinfo {volume} {42}},\ \bibinfo
  {pages} {2282} (\bibinfo {year} {1990})}\BibitemShut {NoStop}%
\bibitem [{\citenamefont {Chen}\ and\ \citenamefont {Tremblay}(1992)}]{Chen92}%
  \BibitemOpen
  \bibfield  {author} {\bibinfo {author} {\bibfnamefont {L.}~\bibnamefont
  {Chen}}\ and\ \bibinfo {author} {\bibfnamefont {A.-M.}\ \bibnamefont
  {Tremblay}},\ }\bibfield  {title} {\bibinfo {title} {Determinant {Monte
  Carlo} for the {Hubbard model} with arbitrarily gauged auxiliary fields},\
  }\href {https://doi.org/10.1142/S0217979292000323} {\bibfield  {journal}
  {\bibinfo  {journal} {Int. J. Mod. Phys. B}\ }\textbf {\bibinfo {volume}
  {06}},\ \bibinfo {pages} {547} (\bibinfo {year} {1992})}\BibitemShut
  {NoStop}%
\bibitem [{\citenamefont {Batrouni}\ and\ \citenamefont {{de
  Forcrand}}(1993)}]{Batrouni93}%
  \BibitemOpen
  \bibfield  {author} {\bibinfo {author} {\bibfnamefont {G.~G.}\ \bibnamefont
  {Batrouni}}\ and\ \bibinfo {author} {\bibfnamefont {P.}~\bibnamefont {{de
  Forcrand}}},\ }\bibfield  {title} {\bibinfo {title} {Fermion sign problem:
  {{Decoupling}} transformation and simulation algorithm},\ }\href
  {https://doi.org/10.1103/PhysRevB.48.589} {\bibfield  {journal} {\bibinfo
  {journal} {Phys. Rev. B}\ }\textbf {\bibinfo {volume} {48}},\ \bibinfo
  {pages} {589} (\bibinfo {year} {1993})}\BibitemShut {NoStop}%
\bibitem [{\citenamefont {Karakuzu}\ \emph {et~al.}(2018)\citenamefont
  {Karakuzu}, \citenamefont {Seki},\ and\ \citenamefont
  {Sorella}}]{KarakuzuPRB2018}%
  \BibitemOpen
  \bibfield  {author} {\bibinfo {author} {\bibfnamefont {S.}~\bibnamefont
  {Karakuzu}}, \bibinfo {author} {\bibfnamefont {K.}~\bibnamefont {Seki}},\
  and\ \bibinfo {author} {\bibfnamefont {S.}~\bibnamefont {Sorella}},\
  }\bibfield  {title} {\bibinfo {title} {Solution of the sign problem for the
  half-filled {H}ubbard-{H}olstein model},\ }\href
  {https://doi.org/10.1103/PhysRevB.98.201108} {\bibfield  {journal} {\bibinfo
  {journal} {Phys. Rev. B}\ }\textbf {\bibinfo {volume} {98}},\ \bibinfo
  {pages} {201108} (\bibinfo {year} {2018})}\BibitemShut {NoStop}%
\bibitem [{\citenamefont {Loh}\ \emph {et~al.}(1990)\citenamefont {Loh},
  \citenamefont {Gubernatis}, \citenamefont {Scalettar}, \citenamefont {White},
  \citenamefont {Scalapino},\ and\ \citenamefont {Sugar}}]{Loh90}%
  \BibitemOpen
  \bibfield  {author} {\bibinfo {author} {\bibfnamefont {E.~Y.}\ \bibnamefont
  {Loh}}, \bibinfo {author} {\bibfnamefont {J.~E.}\ \bibnamefont {Gubernatis}},
  \bibinfo {author} {\bibfnamefont {R.~T.}\ \bibnamefont {Scalettar}}, \bibinfo
  {author} {\bibfnamefont {S.~R.}\ \bibnamefont {White}}, \bibinfo {author}
  {\bibfnamefont {D.~J.}\ \bibnamefont {Scalapino}},\ and\ \bibinfo {author}
  {\bibfnamefont {R.~L.}\ \bibnamefont {Sugar}},\ }\bibfield  {title} {\bibinfo
  {title} {Sign problem in the numerical simulation of many-electron systems},\
  }\href {https://doi.org/10.1103/PhysRevB.41.9301} {\bibfield  {journal}
  {\bibinfo  {journal} {Phys. Rev. B}\ }\textbf {\bibinfo {volume} {41}},\
  \bibinfo {pages} {9301} (\bibinfo {year} {1990})}\BibitemShut {NoStop}%
\bibitem [{\citenamefont {Duane}\ \emph {et~al.}(1987)\citenamefont {Duane},
  \citenamefont {Kennedy}, \citenamefont {Pendleton},\ and\ \citenamefont
  {Roweth}}]{Duane87}%
  \BibitemOpen
  \bibfield  {author} {\bibinfo {author} {\bibfnamefont {S.}~\bibnamefont
  {Duane}}, \bibinfo {author} {\bibfnamefont {A.}~\bibnamefont {Kennedy}},
  \bibinfo {author} {\bibfnamefont {B.~J.}\ \bibnamefont {Pendleton}},\ and\
  \bibinfo {author} {\bibfnamefont {D.}~\bibnamefont {Roweth}},\ }\bibfield
  {title} {\bibinfo {title} {Hybrid {M}onte {C}arlo},\ }\href
  {https://doi.org/https://doi.org/10.1016/0370-2693(87)91197-X} {\bibfield
  {journal} {\bibinfo  {journal} {Physics Letters B}\ }\textbf {\bibinfo
  {volume} {195}},\ \bibinfo {pages} {216} (\bibinfo {year}
  {1987})}\BibitemShut {NoStop}%
\bibitem [{\citenamefont {Neal}(1999)}]{Neal99}%
  \BibitemOpen
  \bibfield  {author} {\bibinfo {author} {\bibfnamefont {R.~M.}\ \bibnamefont
  {Neal}},\ }\href {https://www.cs.toronto.edu/~radford/ftp/jsm99.pdf}
  {\bibinfo {title} {Markov {{Chain Sampling Using Hamiltonian Dynamics}}}}
  (\bibinfo {year} {1999})\BibitemShut {NoStop}%
\bibitem [{\citenamefont {Batrouni}\ \emph {et~al.}(1985)\citenamefont
  {Batrouni}, \citenamefont {Katz}, \citenamefont {Kronfeld}, \citenamefont
  {Lepage}, \citenamefont {Svetitsky},\ and\ \citenamefont
  {Wilson}}]{Batrouni85}%
  \BibitemOpen
  \bibfield  {author} {\bibinfo {author} {\bibfnamefont {G.~G.}\ \bibnamefont
  {Batrouni}}, \bibinfo {author} {\bibfnamefont {G.~R.}\ \bibnamefont {Katz}},
  \bibinfo {author} {\bibfnamefont {A.~S.}\ \bibnamefont {Kronfeld}}, \bibinfo
  {author} {\bibfnamefont {G.~P.}\ \bibnamefont {Lepage}}, \bibinfo {author}
  {\bibfnamefont {B.}~\bibnamefont {Svetitsky}},\ and\ \bibinfo {author}
  {\bibfnamefont {K.~G.}\ \bibnamefont {Wilson}},\ }\bibfield  {title}
  {\bibinfo {title} {Langevin simulations of lattice field theories},\ }\href
  {https://doi.org/10.1103/PhysRevD.32.2736} {\bibfield  {journal} {\bibinfo
  {journal} {Phys. Rev. D}\ }\textbf {\bibinfo {volume} {32}},\ \bibinfo
  {pages} {2736} (\bibinfo {year} {1985})}\BibitemShut {NoStop}%
\bibitem [{\citenamefont {Kennedy}\ and\ \citenamefont
  {Pendleton}(2001)}]{Kennedy01}%
  \BibitemOpen
  \bibfield  {author} {\bibinfo {author} {\bibfnamefont {A.~D.}\ \bibnamefont
  {Kennedy}}\ and\ \bibinfo {author} {\bibfnamefont {B.}~\bibnamefont
  {Pendleton}},\ }\bibfield  {title} {\bibinfo {title} {Cost of the generalised
  hybrid {{Monte Carlo}} algorithm for free field theory},\ }\href
  {https://doi.org/10.1016/S0550-3213(01)00129-8} {\bibfield  {journal}
  {\bibinfo  {journal} {Nuclear Physics B}\ }\textbf {\bibinfo {volume}
  {607}},\ \bibinfo {pages} {456} (\bibinfo {year} {2001})}\BibitemShut
  {NoStop}%
\bibitem [{\citenamefont {Beyl}\ \emph {et~al.}(2018)\citenamefont {Beyl},
  \citenamefont {Goth},\ and\ \citenamefont {Assaad}}]{Beyl18}%
  \BibitemOpen
  \bibfield  {author} {\bibinfo {author} {\bibfnamefont {S.}~\bibnamefont
  {Beyl}}, \bibinfo {author} {\bibfnamefont {F.}~\bibnamefont {Goth}},\ and\
  \bibinfo {author} {\bibfnamefont {F.~F.}\ \bibnamefont {Assaad}},\ }\bibfield
   {title} {\bibinfo {title} {Revisiting the hybrid quantum {{Monte Carlo}}
  method for {{Hubbard}} and electron-phonon models},\ }\href
  {https://doi.org/10.1103/PhysRevB.97.085144} {\bibfield  {journal} {\bibinfo
  {journal} {Phys. Rev. B}\ }\textbf {\bibinfo {volume} {97}},\ \bibinfo
  {pages} {085144} (\bibinfo {year} {2018})}\BibitemShut {NoStop}%
\bibitem [{\citenamefont {Batrouni}\ and\ \citenamefont
  {Scalettar}(2019)}]{Batrouni19}%
  \BibitemOpen
  \bibfield  {author} {\bibinfo {author} {\bibfnamefont {G.~G.}\ \bibnamefont
  {Batrouni}}\ and\ \bibinfo {author} {\bibfnamefont {R.~T.}\ \bibnamefont
  {Scalettar}},\ }\bibfield  {title} {\bibinfo {title} {Langevin simulations of
  a long-range electron-phonon model},\ }\href
  {https://doi.org/10.1103/PhysRevB.99.035114} {\bibfield  {journal} {\bibinfo
  {journal} {Phys. Rev. B}\ }\textbf {\bibinfo {volume} {99}},\ \bibinfo
  {pages} {035114} (\bibinfo {year} {2019})}\BibitemShut {NoStop}%
\bibitem [{\citenamefont {Cohen-Stead}\ \emph
  {et~al.}(2022{\natexlab{a}})\citenamefont {Cohen-Stead}, \citenamefont
  {Bradley}, \citenamefont {Miles}, \citenamefont {Batrouni}, \citenamefont
  {Scalettar},\ and\ \citenamefont {Barros}}]{Cohen-Stead22}%
  \BibitemOpen
  \bibfield  {author} {\bibinfo {author} {\bibfnamefont {B.}~\bibnamefont
  {Cohen-Stead}}, \bibinfo {author} {\bibfnamefont {O.}~\bibnamefont
  {Bradley}}, \bibinfo {author} {\bibfnamefont {C.}~\bibnamefont {Miles}},
  \bibinfo {author} {\bibfnamefont {G.}~\bibnamefont {Batrouni}}, \bibinfo
  {author} {\bibfnamefont {R.}~\bibnamefont {Scalettar}},\ and\ \bibinfo
  {author} {\bibfnamefont {K.}~\bibnamefont {Barros}},\ }\bibfield  {title}
  {\bibinfo {title} {Fast and scalable quantum {M}onte {C}arlo simulations of
  electron-phonon models},\ }\href
  {https://doi.org/10.1103/PhysRevE.105.065302} {\bibfield  {journal} {\bibinfo
   {journal} {Phys. Rev. E}\ }\textbf {\bibinfo {volume} {105}},\ \bibinfo
  {pages} {065302} (\bibinfo {year} {2022}{\natexlab{a}})}\BibitemShut
  {NoStop}%
\bibitem [{\citenamefont {Cohen-Stead}\ \emph
  {et~al.}(2022{\natexlab{b}})\citenamefont {Cohen-Stead}, \citenamefont
  {Barros}, \citenamefont {Scalettar},\ and\ \citenamefont
  {Johnston}}]{CohenSteadPreprint}%
  \BibitemOpen
  \bibfield  {author} {\bibinfo {author} {\bibfnamefont {B.}~\bibnamefont
  {Cohen-Stead}}, \bibinfo {author} {\bibfnamefont {K.}~\bibnamefont {Barros}},
  \bibinfo {author} {\bibfnamefont {R.}~\bibnamefont {Scalettar}},\ and\
  \bibinfo {author} {\bibfnamefont {S.}~\bibnamefont {Johnston}},\ }\bibfield
  {title} {\bibinfo {title} {A hybrid monte carlo study of bond-stretching
  electron-phonon interactions and charge order in the bismuthate family of
  superconductors},\ }\href {https://arxiv.org/abs/2208.02339} {\bibfield
  {journal} {\bibinfo  {journal} {arXiv:2208.02339}\ } (\bibinfo {year}
  {2022}{\natexlab{b}})}\BibitemShut {NoStop}%
\bibitem [{\citenamefont {Lee}(2008)}]{Lee08}%
  \BibitemOpen
  \bibfield  {author} {\bibinfo {author} {\bibfnamefont {D.}~\bibnamefont
  {Lee}},\ }\bibfield  {title} {\bibinfo {title} {Ground state energy at
  unitarity},\ }\href {https://doi.org/10.1103/PhysRevC.78.024001} {\bibfield
  {journal} {\bibinfo  {journal} {Phys. Rev. C}\ }\textbf {\bibinfo {volume}
  {78}},\ \bibinfo {pages} {024001} (\bibinfo {year} {2008})}\BibitemShut
  {NoStop}%
\bibitem [{\citenamefont {{Wan}}\ \emph {et~al.}(2020)\citenamefont {{Wan}},
  \citenamefont {{Zhang}},\ and\ \citenamefont {{Yao}}}]{Wan2020}%
  \BibitemOpen
  \bibfield  {author} {\bibinfo {author} {\bibfnamefont {Z.-Q.}\ \bibnamefont
  {{Wan}}}, \bibinfo {author} {\bibfnamefont {S.-X.}\ \bibnamefont {{Zhang}}},\
  and\ \bibinfo {author} {\bibfnamefont {H.}~\bibnamefont {{Yao}}},\ }\bibfield
   {title} {\bibinfo {title} {{Mitigating sign problem by automatic
  differentiation}},\ }\href {https://arxiv.org/abs/2010.01141} {\bibfield
  {journal} {\bibinfo  {journal} {arXiv:2010.01141}\ } (\bibinfo {year}
  {2020})}\BibitemShut {NoStop}%
\bibitem [{Note1()}]{Note1}%
  \BibitemOpen
  \bibinfo {note} {In describing DQMC codes, one frequently sees written
  $e^{\tau (A+B)}\approx e^{\tau A} e^{\tau B}$, but typically this originates
  from symmetric operator splitting and application of the cyclic property of
  the trace.}\BibitemShut {Stop}%
\bibitem [{\citenamefont {Loh~Jr}\ and\ \citenamefont
  {Gubernatis}(1992)}]{Loh1992}%
  \BibitemOpen
  \bibfield  {author} {\bibinfo {author} {\bibfnamefont {E.}~\bibnamefont
  {Loh~Jr}}\ and\ \bibinfo {author} {\bibfnamefont {J.}~\bibnamefont
  {Gubernatis}},\ }\bibfield  {title} {\bibinfo {title} {Stable numerical
  simulations of models of interacting electrons in condensed matter physics},\
  }\href {https://doi.org/10.1016/B978-0-444-88885-3.50009-3} {\bibfield
  {journal} {\bibinfo  {journal} {Electronic Phase Transitions}\ }\textbf
  {\bibinfo {volume} {32}},\ \bibinfo {pages} {177} (\bibinfo {year}
  {1992})}\BibitemShut {NoStop}%
\bibitem [{\citenamefont {Loh}\ \emph {et~al.}(2005)\citenamefont {Loh},
  \citenamefont {Gubernatis}, \citenamefont {Scalettar}, \citenamefont {White},
  \citenamefont {Scalapino},\ and\ \citenamefont {Sugar}}]{Loh2005}%
  \BibitemOpen
  \bibfield  {author} {\bibinfo {author} {\bibfnamefont {E.}~\bibnamefont
  {Loh}}, \bibinfo {author} {\bibfnamefont {J.}~\bibnamefont {Gubernatis}},
  \bibinfo {author} {\bibfnamefont {R.}~\bibnamefont {Scalettar}}, \bibinfo
  {author} {\bibfnamefont {S.}~\bibnamefont {White}}, \bibinfo {author}
  {\bibfnamefont {D.}~\bibnamefont {Scalapino}},\ and\ \bibinfo {author}
  {\bibfnamefont {R.}~\bibnamefont {Sugar}},\ }\bibfield  {title} {\bibinfo
  {title} {Numerical stability and the sign problem in the determinant quantum
  {M}onte {C}arlo method},\ }\href {https://doi.org/10.1142/S0129183105007911}
  {\bibfield  {journal} {\bibinfo  {journal} {International Journal of Modern
  Physics C}\ }\textbf {\bibinfo {volume} {16}},\ \bibinfo {pages} {1319}
  (\bibinfo {year} {2005})}\BibitemShut {NoStop}%
\bibitem [{\citenamefont {Bauer}(2020)}]{Bauer2020}%
  \BibitemOpen
  \bibfield  {author} {\bibinfo {author} {\bibfnamefont {C.}~\bibnamefont
  {Bauer}},\ }\bibfield  {title} {\bibinfo {title} {Fast and stable determinant
  quantum {M}onte {C}arlo},\ }\href
  {https://scipost.org/10.21468/SciPostPhysCore.2.2.011} {\bibfield  {journal}
  {\bibinfo  {journal} {SciPost Physics Core}\ }\textbf {\bibinfo {volume}
  {2}},\ \bibinfo {pages} {011} (\bibinfo {year} {2020})}\BibitemShut {NoStop}%
\bibitem [{\citenamefont {Kennedy}(1990)}]{Kennedy90}%
  \BibitemOpen
  \bibfield  {author} {\bibinfo {author} {\bibfnamefont {A.}~\bibnamefont
  {Kennedy}},\ }\bibfield  {title} {\bibinfo {title} {The theory of hybrid
  stochastic algorithms},\ }in\ \href@noop {} {\emph {\bibinfo {booktitle}
  {Probabilistic methods in quantum field theory and quantum gravity}}}\
  (\bibinfo  {publisher} {Springer},\ \bibinfo {year} {1990})\ pp.\ \bibinfo
  {pages} {209--223}\BibitemShut {NoStop}%
\bibitem [{\citenamefont {Besag}(1994)}]{Besag1994}%
  \BibitemOpen
  \bibfield  {author} {\bibinfo {author} {\bibfnamefont {J.}~\bibnamefont
  {Besag}},\ }\bibfield  {title} {\bibinfo {title} {Comments on
  ``{R}epresentations of knowledge in complex systems'' by {U}. {G}renander and
  {M}. {I}. {M}iller},\ }\href@noop {} {\bibfield  {journal} {\bibinfo
  {journal} {J. Roy. Statist. Soc. Ser. B}\ }\textbf {\bibinfo {volume} {56}},\
  \bibinfo {pages} {4} (\bibinfo {year} {1994})}\BibitemShut {NoStop}%
\bibitem [{\citenamefont {Troyer}\ and\ \citenamefont
  {Wiese}(2005)}]{Troyer05}%
  \BibitemOpen
  \bibfield  {author} {\bibinfo {author} {\bibfnamefont {M.}~\bibnamefont
  {Troyer}}\ and\ \bibinfo {author} {\bibfnamefont {U.-J.}\ \bibnamefont
  {Wiese}},\ }\bibfield  {title} {\bibinfo {title} {Computational complexity
  and fundamental limitations to fermionic quantum {M}onte {C}arlo
  simulations},\ }\href {https://doi.org/10.1103/PhysRevLett.94.170201}
  {\bibfield  {journal} {\bibinfo  {journal} {Phys. Rev. Lett.}\ }\textbf
  {\bibinfo {volume} {94}},\ \bibinfo {pages} {170201} (\bibinfo {year}
  {2005})}\BibitemShut {NoStop}%
\bibitem [{\citenamefont {Iglovikov}\ \emph {et~al.}(2015)\citenamefont
  {Iglovikov}, \citenamefont {Khatami},\ and\ \citenamefont
  {Scalettar}}]{IglovikovPRB2015}%
  \BibitemOpen
  \bibfield  {author} {\bibinfo {author} {\bibfnamefont {V.~I.}\ \bibnamefont
  {Iglovikov}}, \bibinfo {author} {\bibfnamefont {E.}~\bibnamefont {Khatami}},\
  and\ \bibinfo {author} {\bibfnamefont {R.~T.}\ \bibnamefont {Scalettar}},\
  }\bibfield  {title} {\bibinfo {title} {Geometry dependence of the sign
  problem in quantum {Monte Carlo} simulations},\ }\href
  {https://doi.org/10.1103/PhysRevB.92.045110} {\bibfield  {journal} {\bibinfo
  {journal} {Phys. Rev. B}\ }\textbf {\bibinfo {volume} {92}},\ \bibinfo
  {pages} {045110} (\bibinfo {year} {2015})}\BibitemShut {NoStop}%
\bibitem [{\citenamefont {Iazzi}\ \emph {et~al.}(2016)\citenamefont {Iazzi},
  \citenamefont {Soluyanov},\ and\ \citenamefont {Troyer}}]{Iazzi16}%
  \BibitemOpen
  \bibfield  {author} {\bibinfo {author} {\bibfnamefont {M.}~\bibnamefont
  {Iazzi}}, \bibinfo {author} {\bibfnamefont {A.~A.}\ \bibnamefont
  {Soluyanov}},\ and\ \bibinfo {author} {\bibfnamefont {M.}~\bibnamefont
  {Troyer}},\ }\bibfield  {title} {\bibinfo {title} {Topological origin of the
  fermion sign problem},\ }\href {https://doi.org/10.1103/PhysRevB.93.115102}
  {\bibfield  {journal} {\bibinfo  {journal} {Phys. Rev. B}\ }\textbf {\bibinfo
  {volume} {93}},\ \bibinfo {pages} {115102} (\bibinfo {year}
  {2016})}\BibitemShut {NoStop}%
\bibitem [{\citenamefont {Mondaini}\ \emph {et~al.}(2022)\citenamefont
  {Mondaini}, \citenamefont {Tarat},\ and\ \citenamefont
  {Scalettar}}]{mondaini22}%
  \BibitemOpen
  \bibfield  {author} {\bibinfo {author} {\bibfnamefont {R.}~\bibnamefont
  {Mondaini}}, \bibinfo {author} {\bibfnamefont {S.}~\bibnamefont {Tarat}},\
  and\ \bibinfo {author} {\bibfnamefont {R.~T.}\ \bibnamefont {Scalettar}},\
  }\bibfield  {title} {\bibinfo {title} {Quantum critical points and the sign
  problem},\ }\href {https://www.science.org/doi/10.1126/science.abg9299}
  {\bibfield  {journal} {\bibinfo  {journal} {Science}\ }\textbf {\bibinfo
  {volume} {375}},\ \bibinfo {pages} {418} (\bibinfo {year}
  {2022})}\BibitemShut {NoStop}%
\bibitem [{\citenamefont {Li}\ \emph {et~al.}(2018)\citenamefont {Li},
  \citenamefont {Tang}, \citenamefont {Maier},\ and\ \citenamefont
  {Johnston}}]{LiPRB2018}%
  \BibitemOpen
  \bibfield  {author} {\bibinfo {author} {\bibfnamefont {S.}~\bibnamefont
  {Li}}, \bibinfo {author} {\bibfnamefont {Y.}~\bibnamefont {Tang}}, \bibinfo
  {author} {\bibfnamefont {T.~A.}\ \bibnamefont {Maier}},\ and\ \bibinfo
  {author} {\bibfnamefont {S.}~\bibnamefont {Johnston}},\ }\bibfield  {title}
  {\bibinfo {title} {Phase competition in a one-dimensional three-orbital
  {H}ubbard-{H}olstein model},\ }\href
  {https://doi.org/10.1103/PhysRevB.97.195116} {\bibfield  {journal} {\bibinfo
  {journal} {Phys. Rev. B}\ }\textbf {\bibinfo {volume} {97}},\ \bibinfo
  {pages} {195116} (\bibinfo {year} {2018})}\BibitemShut {NoStop}%
\bibitem [{\citenamefont {Kung}\ \emph {et~al.}(2016)\citenamefont {Kung},
  \citenamefont {Chen}, \citenamefont {Wang}, \citenamefont {Huang},
  \citenamefont {Nowadnick}, \citenamefont {Moritz}, \citenamefont {Scalettar},
  \citenamefont {Johnston},\ and\ \citenamefont {Devereaux}}]{KungPRB2016}%
  \BibitemOpen
  \bibfield  {author} {\bibinfo {author} {\bibfnamefont {Y.~F.}\ \bibnamefont
  {Kung}}, \bibinfo {author} {\bibfnamefont {C.-C.}\ \bibnamefont {Chen}},
  \bibinfo {author} {\bibfnamefont {Y.}~\bibnamefont {Wang}}, \bibinfo {author}
  {\bibfnamefont {E.~W.}\ \bibnamefont {Huang}}, \bibinfo {author}
  {\bibfnamefont {E.~A.}\ \bibnamefont {Nowadnick}}, \bibinfo {author}
  {\bibfnamefont {B.}~\bibnamefont {Moritz}}, \bibinfo {author} {\bibfnamefont
  {R.~T.}\ \bibnamefont {Scalettar}}, \bibinfo {author} {\bibfnamefont
  {S.}~\bibnamefont {Johnston}},\ and\ \bibinfo {author} {\bibfnamefont
  {T.~P.}\ \bibnamefont {Devereaux}},\ }\bibfield  {title} {\bibinfo {title}
  {Characterizing the three-orbital {Hubbard} model with determinant quantum
  {Monte Carlo}},\ }\href {https://doi.org/10.1103/PhysRevB.93.155166}
  {\bibfield  {journal} {\bibinfo  {journal} {Phys. Rev. B}\ }\textbf {\bibinfo
  {volume} {93}},\ \bibinfo {pages} {155166} (\bibinfo {year}
  {2016})}\BibitemShut {NoStop}%
\bibitem [{\citenamefont {Mou}\ \emph {et~al.}(2022)\citenamefont {Mou},
  \citenamefont {Mondaini},\ and\ \citenamefont {Scalettar}}]{Mou2022}%
  \BibitemOpen
  \bibfield  {author} {\bibinfo {author} {\bibfnamefont {Y.}~\bibnamefont
  {Mou}}, \bibinfo {author} {\bibfnamefont {R.}~\bibnamefont {Mondaini}},\ and\
  \bibinfo {author} {\bibfnamefont {R.~T.}\ \bibnamefont {Scalettar}},\
  }\bibfield  {title} {\bibinfo {title} {Bilayer hubbard model: Analysis based
  on the fermionic sign problem},\ }\href
  {https://doi.org/10.1103/PhysRevB.106.125116} {\bibfield  {journal} {\bibinfo
   {journal} {Phys. Rev. B}\ }\textbf {\bibinfo {volume} {106}},\ \bibinfo
  {pages} {125116} (\bibinfo {year} {2022})}\BibitemShut {NoStop}%
\bibitem [{\citenamefont {Johnston}\ \emph {et~al.}(2013)\citenamefont
  {Johnston}, \citenamefont {Nowadnick}, \citenamefont {Kung}, \citenamefont
  {Moritz}, \citenamefont {Scalettar},\ and\ \citenamefont
  {Devereaux}}]{JohnstonPRB2013}%
  \BibitemOpen
  \bibfield  {author} {\bibinfo {author} {\bibfnamefont {S.}~\bibnamefont
  {Johnston}}, \bibinfo {author} {\bibfnamefont {E.~A.}\ \bibnamefont
  {Nowadnick}}, \bibinfo {author} {\bibfnamefont {Y.~F.}\ \bibnamefont {Kung}},
  \bibinfo {author} {\bibfnamefont {B.}~\bibnamefont {Moritz}}, \bibinfo
  {author} {\bibfnamefont {R.~T.}\ \bibnamefont {Scalettar}},\ and\ \bibinfo
  {author} {\bibfnamefont {T.~P.}\ \bibnamefont {Devereaux}},\ }\bibfield
  {title} {\bibinfo {title} {Determinant quantum {Monte Carlo} study of the
  two-dimensional single-band {Hubbard-Holstein} model},\ }\href
  {https://doi.org/10.1103/PhysRevB.87.235133} {\bibfield  {journal} {\bibinfo
  {journal} {Phys. Rev. B}\ }\textbf {\bibinfo {volume} {87}},\ \bibinfo
  {pages} {235133} (\bibinfo {year} {2013})}\BibitemShut {NoStop}%
\bibitem [{\citenamefont {Bai}\ \emph {et~al.}(2009)\citenamefont {Bai},
  \citenamefont {Chen}, \citenamefont {Scalettar},\ and\ \citenamefont
  {Yamazaki}}]{Bai09}%
  \BibitemOpen
  \bibfield  {author} {\bibinfo {author} {\bibfnamefont {Z.}~\bibnamefont
  {Bai}}, \bibinfo {author} {\bibfnamefont {W.}~\bibnamefont {Chen}}, \bibinfo
  {author} {\bibfnamefont {R.}~\bibnamefont {Scalettar}},\ and\ \bibinfo
  {author} {\bibfnamefont {I.}~\bibnamefont {Yamazaki}},\ }\bibfield  {title}
  {\bibinfo {title} {Numerical methods for quantum {M}onte {C}arlo simulations
  of the {H}ubbard model},\ }in\ \href
  {https://doi.org/10.1142/9789814273268_0001} {\emph {\bibinfo {booktitle}
  {Multi-{{Scale Phenomena}} in {{Complex Fluids}}}}}\ (\bibinfo {year}
  {2009})\ pp.\ \bibinfo {pages} {1--110}\BibitemShut {NoStop}%
\end{thebibliography}%

\end{document}